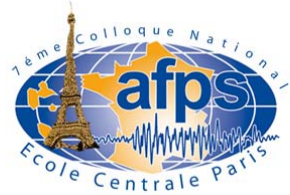

# Analyse de vulnérabilité sismique à grande échelle par utilisation des propriétés dynamiques expérimentales des bâtiments

**Clotaire Michel\* — Philippe Guéguen\*,\*\***

*\* Laboratoire de Géophysique Interne et Tectonophysique (LGIT)*

*CNRS/Université Joseph Fourier*

*1381 rue de la Piscine, 38041 Grenoble cedex 9*

*cmichel@obs.ujf-grenoble.fr*

*\*\* Laboratoire Central des Ponts-et-Chaussées*

*58 boulevard Lefebvre, 75732 Paris cedex 15*

RÉSUMÉ. *Il existe deux familles de méthodes d'analyse de la vulnérabilité sismique : les méthodes de vulnérabilité observée et de vulnérabilité calculée. Dans les premières, on utilise le retour d'expérience des séismes passés pour estimer la vulnérabilité des bâtiments représentés par leurs caractéristiques structurales. Pour les secondes, la vulnérabilité du bâtiment est calculée sur la base des informations structurales disponibles à l'aide de modèles numériques plus ou moins sophistiqués. Dans les 2 cas se pose le problème de la connaissance du bâti existant qui n'est jamais parfaite. Nous proposons une nouvelle méthode d'estimation de la vulnérabilité fondée sur les caractéristiques modales expérimentales (fréquences de résonance, déformée modale et amortissement) obtenues sous vibrations ambiantes qui conduisent à une modélisation analytique simplifiée du bâtiment dans sa partie élastique. La sollicitation par un panel de séismes permet de déterminer les mouvements du sol qui conduisent au premier niveau d'endommagement et donc d'estimer sa vulnérabilité. Le paramètre utilisé pour déterminer le premier niveau d'endommagement est la déformation inter-étage dont les valeurs limites sont données par la méthode américaine HAZUS. Cette méthode est appliquée à la ville de Grenoble où 60 bâtiments ont été instrumentés.*

ABSTRACT. *Two different way of assessing seismic vulnerability are available nowadays: observed or empirical and calculated vulnerability assessment methods. The first methods are based on observed damage after earthquakes correlated with the structural properties of buildings, whereas the second methods are based on numerical models more or less representing the buildings. In both cases, the trouble is the imperfect knowledge of existing buildings. We propose here a new method for estimating the vulnerability based on experimental modal parameters (resonance frequencies, modal shapes and damping ratio) estimated under ambient vibrations. They allow to build up a simplified numerical model of the elastic building behaviour. The motion produced by numerous earthquakes leads to determine its first damage level and therefore its vulnerability. An inter-story drift threshold based on HAZUS values defines the first damage level of the building. This method is applied to the Grenoble (France) city in which 60 buildings have been instrumented.*

MOTS-CLÉS : *Vulnérabilité sismique, bâti existant, analyse modale, vibrations ambiantes, courbes de fragilité.*

**1. Introduction**

Analyser la vulnérabilité du bâti existant est un exercice difficile. Deux familles de méthodes sont généralement utilisées : d'une part, les méthodes empiriques, fondée sur le retour d'expérience et sur les caractéristiques structurales sommaires des bâtiments, sont utilisées à grande échelle (ville, région) (HAZUS , 1999 ; Risk-UE, 2003). D'autre part, pour un nombre plus restreint de bâtiments, les méthodes de vulnérabilité calculée utilisent la modélisation plus ou moins sophistiquée du bâtiment considéré pour obtenir sa courbe de capacité reliant forces et déplacements (méthode du Push-over) ou un modèle numérique complet. La combinaison avec un aléa (déterministe ou probabiliste) permet d'estimer l'endommagement de la structure. Plus ou moins coûteuses, toutes ces méthodes doivent se contenter d'une connaissance très partielle des bâtiments étudiés, les plans, les dispositions constructives et l'effet du vieillissement étant souvent inconnus. C'est généralement le domaine de compétence des experts qui ont une bonne pratique des structures et qui procèdent par analogie. Cependant, dans les pays à sismicité modérée comme la France, il n'existe pas de retour d'expérience de séismes destructeurs significatifs. C'est un problème à surmonter, d'une part, pour les méthodes empiriques qui se doivent d'adapter des méthodes fondées sur un contexte de bâti différent et d'autre part pour les méthodes calculées car les experts ne connaissent pas la tenue des structures françaises aux séismes. Les méthodes calculées ont donc été parfois calibrées à l'aide de modèles réduits excités sous table vibrante, ce qui nécessite de lourds investissements. Compte tenu de l'hétérogénéité des structures existantes, il faut pouvoir obtenir un grand nombre de données expérimentales pour obtenir une méthode fiable. Cela n'est permis que par les méthodes non destructives.

Nous proposons ici d'utiliser les enregistrements de vibrations ambiantes pour déterminer les paramètres modaux (fréquences de résonance, déformées modales et amortissements) des structures. Ces mesures sont réalisables pour un nombre assez important de bâtiments pour un coût assez faible. Ces paramètres permettent de calibrer un modèle élastique de la structure pour calculer les déplacements en chaque point de celle-ci. Le choix d'une limite élastique doit s'en remettre aux méthodes existantes (empiriques ou calculées) mais il permet de déterminer si un mouvement sismique endommage ou pas une structure. Enfin, la combinaison de plusieurs enregistrements conduit à établir la courbe de fragilité du bâtiment qui relie la probabilité de dépassement de la limite d'endommagement à la taille du séisme. Cette méthodologie est appliquée en fil rouge à la ville de Grenoble.

**2. Enregistrements de vibrations ambiantes et analyse modale**

Les structures sont excitées en permanence par le bruit de fond sismique (océans, activités humaines…), le vent et des sollicitations internes. La propriété de bruit blanc du bruit de fond sismique permet d'utiliser les enregistrements dans la structure sans en connaître la source grâce aux méthodes d'analyse modale dites « output-only » (Peeters, 2000). L'enregistrement dans une structure avec un point à chaque étage prends environ une demi-journée pour 2 opérateurs. Pour cela, nous avons utilisé une station d'acquisition Cityshark II (Châtelain et al., 2000) qui permet l'acquisition de 18 voies simultanément et 6 capteurs vélocimétriques Lennartz 3D 5s avec une réponse plate en vitesse entre 0.2 et 50 Hz. Nous avons enregistré les vibrations ambiantes de 60 bâtiments de types variés (béton et maçonnerie) de Grenoble avec un point par étage au minimum. Chaque enregistrement a duré 15 min à une fréquence d'échantillonnage de 200 Hz.

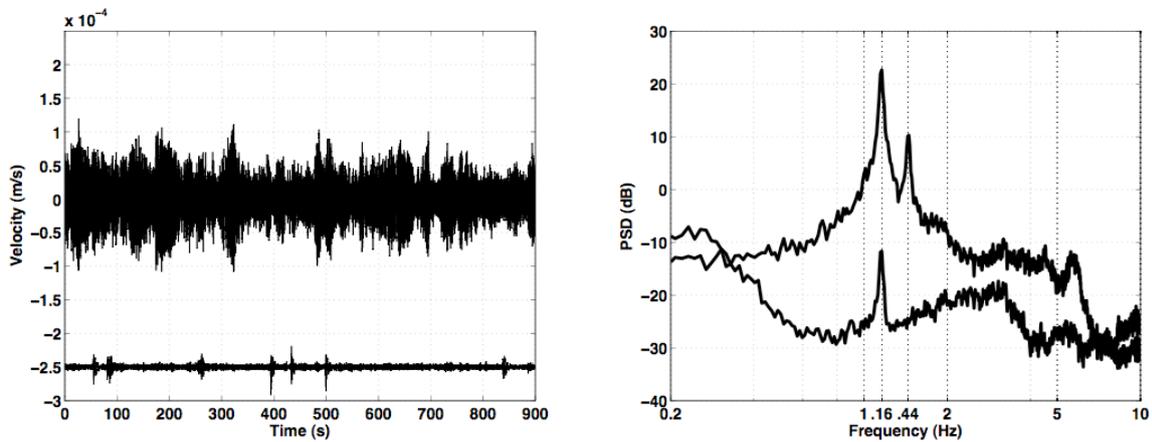

**Figure 1.** *Enregistrement de vibrations ambiantes (composante longitudinale) dans l'hôtel de ville de Grenoble au rez-de-chaussée et au toit et leurs spectres (densité spectrale de puissance).*

Nous avons utilisé la Frequency Domain Decomposition FDD (Brincker et al., 2001a) pour réaliser l'analyse modale de ces données. Cette méthode permet de tirer parti des enregistrements simultanés. En outre, elle permet de séparer les modes même s'ils sont proches. Il s'agit de réaliser une décomposition en valeurs singulières des matrices de densité spectrales, c'est-à-dire des transformées de Fourier des matrices de corrélation entre tous les enregistrements simultanés. La première valeur singulière présente des pics aux fréquences de résonance dont la déformée modale est le vecteur singulier correspondant. À l'aide d'un critère de similarité des déformées modales (critère MAC) avec la déformée du pic comme référence, il est possible d'isoler la cloche correspondant à chaque mode (qui se prolonge sur les valeurs singulières suivantes). À partir de cette cloche, on obtient une valeur affinée de la fréquence et une valeur d'amortissement par transformée de Fourier inverse et calcul du décrément logarithmique (Brincker et al., 2001b). L'estimation des premières fréquences propres, déformées et amortissements est fiable si la durée d'enregistrement est assez longue. Le coefficient de participation diminue fortement avec l'ordre du mode donc l'importance des modes d'ordre élevé dans la réponse sismique est, la plupart du temps, négligeable.

Les fréquences propres ont tendance à diminuer avec l'augmentation de l'amplitude de la sollicitation par mise en fonctionnement de fissures dans le béton (Mazars, communication personnelle). Cependant, contrairement aux mesures réalisées sur des modèles réduits, de nombreux auteurs ont montré que cette diminution était très faible dans les bâtiments réels, des vibrations ambiantes ($10^{-4}$g) aux séismes modérés (0.1g) (Dunand et al., 2006 ; Hans et al., 2005 ; Michel et Guéguen, 2007). Au delà de la limite élastique, les fréquences propres chutent de manière permanente en raison de l'endommagement de la structure. Par ailleurs, les déformées modales sont assez stables tant que les dommages sont faibles. En ce qui concerne l'amortissement, il devrait en théorie augmenter lorsque la sollicitation augmente mais aucune tendance claire et simple n'a pu être mise en évidence jusqu'ici (Dunand, 2005), les valeurs obtenues sont donc à prendre avec la plus grande prudence.

## 3. Modélisation simple

Les paramètres modaux obtenus sous vibrations ambiantes sont ensuite utilisés pour calibrer un modèle de bâtiment. Comme il s'agit ici de réaliser un modèle pour de nombreux bâtiments, le modèle choisi est un modèle brochette (masses concentrées aux étages) à une dimension. Pour des applications plus spécifiques, un modèle plus complexe 3D en éléments finis par exemple peut être calibré grâce aux paramètres modaux (« model updating ») . Le modèle brochette 1D est bien adapté à la géométrie des enregistrements, généralement réalisés dans la cage d'escalier. Le mouvement élastique de la structure à chaque étage {U(t)} est donné analytiquement par l'intégrale de Duhamel (Clough and Penzien, 1993) si on suppose une masse constante pour chaque étage [M], connaissant les premiers modes de vibration ([Φ] les déformées modales, {ω} les fréquences et {ξ} les amortissements) et le mouvement du sol $U_s(t)$:

$$\{U(t)\} = [\Phi]\{y(t)\} + U_s(t) \qquad [1]$$

$$\text{avec } \forall j \in [1,N] \quad y_j(t) = \frac{-p_j}{\omega'} \int_0^t U_s''(\tau) e^{-\xi_j \omega_j (t-\tau)} \sin(\omega'(t-\tau)) d\tau \quad [2]$$

$$\omega_j'^2 = \omega_j^2(1-\xi_j^2) \text{ et } p_j = \frac{\{\Phi_j\}^T[M]\{1\}}{\{\Phi_j\}^T[M]\{\Phi_j\}} = \frac{\sum_{i=1}^N \Phi_{ij}}{\sum_{i=1}^N \Phi_{ij}^2} \text{ le facteur de participation du mode j.}$$

Pour un scénario de séisme ($U_s(t)$ donné), on obtient donc le déplacement de chaque étage.

**4. Limite élastique et courbes de fragilité**

Le modèle issu des paramètres modaux expérimentaux donne donc le mouvement de la structure sous séisme. Le génie parasismique utilise actuellement plus volontiers les déplacements pour décrire le mouvement des structures. Les seuils d'endommagement, plus particulièrement, sont donnés dans les méthodes simplifiées, comme HAZUS, en déformation inter-étage. Nous avons donc opté pour une limite élastique en déformation inter-étage (« inter-story drift »), déplacement inter-étage ramené à la hauteur de l'étage. Les valeurs de limite en déformation de différentes provenances ont été compilées par la FEMA (2003) pour différents types de bâtiments. Bien que les types de bâtiments en France soient assez différents des bâtiments américains, nous avons tout même retenu les valeurs données pour le niveau de dommage « léger ». Il varie de $10^{-3}$ pour la maçonnerie à $4 \times 10^{-3}$ pour certains bâtiments en béton. L'utilisation de ce paramètre suppose que c'est la composante en cisaillement qui endommage la structure, ce qui n'est plus vrai pour des états d'endommagements avancés pour la plupart des types de bâtiments, en particulier en maçonnerie. Cette méthode est donc adaptée pour déterminer l'endommagement attendu d'une ville dans le cas d'un séisme modéré pour lequel des ruines ne sont pas à attendre. Elle peut également servir de point de départ élastique à des calculs en plasticité.

Nous avons sélectionné pour cette étude 150 séismes de magnitude 4.5, 5 et 5.5 simulés sur la faille bordière de Belledonne (Thouvenot et al., 2003) à 15 km de Grenoble par Causse et al. (2006) à l'aide de fonctions de Green empiriques calculées au centre du bassin. Ils représentent la variabilité du mouvement sismique attendu pour ce scénario, qui n'est pas le seul scénario possible bien qu'il soit souvent évoqué comme dominant l'aléa local.

Nous avons choisi de représenter le mouvement du sol par la valeur de son spectre de réponse en déplacement à la fréquence de la structure à ξ% (ξ l'amortissement déterminé par les enregistrements ou 5% s'il n'a pu être déterminé) noté par la suite Sd (en m). Ce choix implique que les paramètres qui vont avoir une influence sur la courbe de fragilité sont : le type de la structure (valeur de déformation inter-étage limite), sa déformée modale (régularité, coefficient de participation) et sa hauteur. On suppose, à l'aide du théorème central limite, que la distribution des déformations inter-étage maximales suit une loi log-normale. Pour chaque classe de Sd, on estime donc les paramètres de cette loi puis on détermine la probabilité de dépasser la déformation limite. Cette probabilité de dépassement constitue un point de la courbe de fragilité. En supposant qu'elle constitue elle-même la densité de probabilité cumulée d'une loi log-normale, on détermine les paramètres $\ln(\mu)$ et $\sigma$ liés à cette loi. L'exponentielle $\mu$ du premier paramètre donne le Sd qui aurait une probabilité de 0.5 d'endommager la structure (médiane). L'écart type $\sigma$ donne la variabilité liée au mouvement sismique en supposant que les déformées sont fiables. La loi de probabilité cumulée s'écrit à l'aide la fonction d'erreur (*erf*):

$$P[d > "Slight"](Sd) = \frac{1}{2} + \frac{1}{2} erf\left(\frac{\ln(Sd) - \ln(\mu)}{\sigma\sqrt{2}}\right) \qquad [3]$$

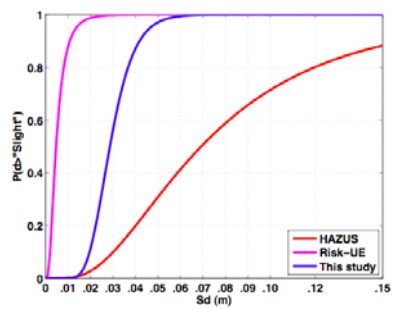

**Figure 2.** *Courbes de fragilité de l'hôtel de ville de Grenoble représentant la probabilité de premier endommagement du bâtiment en fonction du spectre de réponse en déplacement à la fréquence de la structure. La plus pessimiste (à gauche) représente le type RC2H PreCode de Risk-UE (2003) correspondant à ce bâtiment, la plus optimiste (à droite) a été calculée à partir du document HAZUS (FEMA, 2003). Celle du milieu est le résultat de la méthode décrite ici.*

La courbe de fragilité obtenue de cette manière pour l'hôtel de ville de Grenoble est comparée, sur la figure 2, avec les courbes données par Risk-UE (2003) et HAZUS (2003) correspondant au type de cette structure. La courbe Risk-UE, donnée par l'Université de Bucarest (UTCB), semble particulièrement pessimiste alors que la courbe HAZUS correspond à notre calcul à la différence près que c'est un bâtiment parfaitement régulier qui est supposé. La différence est assez importante en terme de vulnérabilité comme on peut le constater figure 2. Pour l'application aux bâtiments testés, la courbe de fragilité n'a pas pu être déterminée pour tous à cause du choix des séismes de sollicitation. En effet, pour certains bâtiments, pas assez de séismes atteignent la limite élastique pour déterminer la loi de fragilité. Les résultats sont présentés à la figure 3. La maçonnerie est logiquement plus vulnérable que le béton avec une probabilité d'endommagement de 0.5 pour des Sd de 1cm environ. Cette valeur est seulement un peu supérieure pour les bâtiments en béton d'avant 1950 (BA1 et BA2) et des années 1960 (BA4), alors que les bâtiments plus récents (années 1970 BA5 et après BA6) sont beaucoup moins vulnérables. Le plus « résistant » d'après ces courbes est le bâtiment 16, la tour Mont Blanc, l'une des tours de l'Ile Verte (28 étages) avec un Sd de probabilité d'endommagement 0.5 de 6cm.

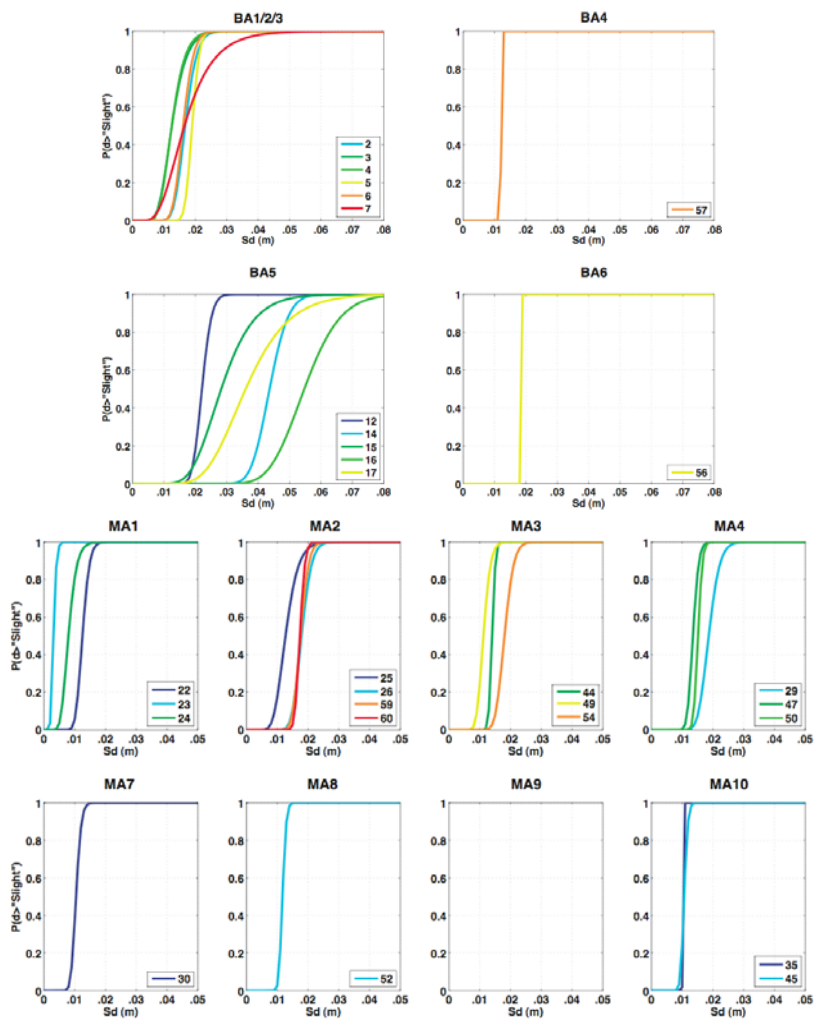

**Figure 3.** *Courbes de fragilité pour les bâtiments de Grenoble en béton (en haut) et en maçonnerie (en bas) représentant la probabilité d'endommagement pour un séisme avec un déplacement spectral Sd à la fréquence de la structure. Chaque graphique représente un type de bâtiments de la typologie grenobloise.*

### 5. Probabilité d'endommagement

Connaissant la fréquence de la structure et son amortissement, on peut déterminer pour un aléa donné (séisme scénario, spectre réglementaire, …) la probabilité d'endommagement de la structure sur la courbe de fragilité. On rappelle qu'il s'agit du premier endommagement ou « seuil d'intégrité » comme définit par Boutin et al. (2005). Cette valeur est particulièrement utile pour les pays à sismicité modérée, comme la France, pour lesquels il est plus probable d'avoir des mouvements sismiques provoquant des dommages légers que la ruine de bâtiments. Déterminer quelles structures ont le plus de chance d'être touchés par ce type de dommage (qui peut blesser des personnes) permet de hiérarchiser les priorités de réhabilitation, en particulier pour limiter les chutes d'éléments non-structuraux et de déterminer les quartiers de la ville qui seraient les plus touchés en cas de séisme.

Nous avons choisi d'utiliser comme aléa la sollicitation de dimensionnement préconisée dans les codes actuels. Cela permet de comparer la performance des bâtiments existants aux bâtiments récents construits selon les normes. Nous avons utilisé la valeur d'accélération du nouveau zonage de la France fixée à $a_g$=1.6 m/s² ainsi que le spectre de dimensionnement de l'Eurocode 8. Il faut noter qu'il est moins défavorable que le spectre PS92 compte tenu du coefficient appliqué pendant la période de transition entre les deux spectres.

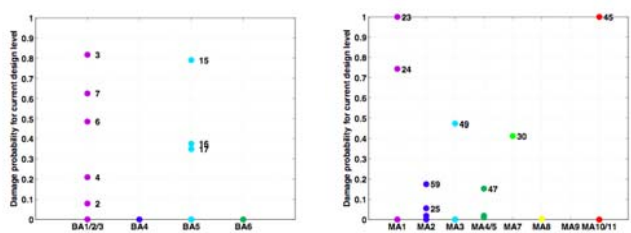

**Figure 4.** *Risque d'endommagement des bâtiments étudiés à Grenoble relativement au niveau de dimensionnement actuel ($a_g$=1.6m/s² ; spectre EC8). À gauche, les bâtiments en béton, à droite les bâtiments en maçonnerie, les abscisses correspondent à un type de bâtiment défini dans la typologie grenobloise, le numéro renvoie au bâtiment de cette typologie.*

Les résultats en terme de risque pour Grenoble sont très riches (figure 4). On ne constate tout d'abord pas d'écart flagrant entre le béton et la maçonnerie, malgré une vulnérabilité de la maçonnerie plus marquée. Les bâtiments en maçonnerie ont, en effet, des fréquences de résonance généralement assez élevée, ce qui implique que leur sollicitation est plus faible dans le spectre de dimensionnement. Il s'agit des résultats élastiques, on ne tient donc évidemment pas compte ici du fait que la réserve plastique des bâtiments en béton est nettement plus importante. Pour le béton, les types BA1, BA2 et BA5 semblent les plus préoccupants. Il s'agit des bâtiments les plus hauts (plus de 7 étages), c'est-à-dire avec la plus basse fréquence (moins de 2 Hz) donc la plus forte sollicitation, qu'ils soient antérieurs à 1950 (BA1, BA2 situés sur les Grands Boulevards) ou datant des années 1970 (BA5). Le bâtiment 15 est l'hôtel de ville de Grenoble (13 étages), le 16 la Tour Mont-blanc, l'une des trois grandes tours de l'Ile Verte (28 étages), très basse fréquence (0.7 Hz) mais moins vulnérable et le 17 la tour ARPEJ (15 étages) sur le campus. Les bâtiments des années 1960 (BA4) et les bâtiments plus récents (BA6), plus bas (4-5 étages au maximum), ne sont pas endommagés. Pour la maçonnerie, les types les plus endommagés sont le type M10, des bâtiments de plus de 6 étages et le type M1, les bâtiments en maçonnerie tout venant composant la vieille ville. Les bâtiments des autres types de maçonnerie ont également une probabilité non nulle d'être endommagés. Le nombre des bâtiments présents sur la figure 4 reste cependant limité et d'autres études sont nécessaires pour donner des tendances à l'échelle de la ville plus fiables.

## 6. Conclusion

Les enregistrements de vibrations ambiantes in situ permettent donc de déterminer le comportement dynamique des structures sous séisme modéré de manière pertinente et sans moyen de calcul. Elles intègrent les problématiques des dispositions constructives, du vieillissement, d'endommagement, etc., souvent impossibles à prendre en compte dans un calcul pour le bâti existant. Elles permettent de mener à la vulnérabilité sismique de la structure décrite par une courbe de fragilité donnant la probabilité de dépasser le « seuil d'intégrité », c'est à dire de déterminer si la structure commence à s'endommager ou pas. La connaissance de ce seuil est particulièrement intéressante dans des pays à sismicité modérée où les mouvements en mesure de produire ce niveau d'endommagement sont plus probables que les séismes destructeurs. Les paramètres déterminés dans le domaine élastique par les enregistrements peuvent ensuite être utilisés comme point de départ à un calcul non-linéaire.

Les courbes de fragilité déterminées pour les bâtiments de la ville de Grenoble sont cohérentes avec ce que l'on sait du bâti : la maçonnerie, en particulier le type MA1 (maçonnerie tout venant), est plus vulnérable que le béton. Les tours de l'Ile Verte, très régulières sont les moins vulnérables. La comparaison avec les méthodes HAZUS et Risk-UE montre que l'utilisation des enregistrements permet d'affiner la courbe de fragilité et que l'hypothèse de régularité prise dans HAZUS est très optimiste pour le bâti français. En convoluant la vulnérabilité avec le niveau de demande réglementaire, on trouve que l'écart entre le béton et la maçonnerie se resserre : les grands bâtiments en béton et en maçonnerie sont en effet plus sollicités en déplacement.

Les évolutions possibles de cette méthode sont nombreuses comme une prise en compte de la torsion et l'amélioration des valeurs de déformation inter-étage limite. Par ailleurs, le couplage avec un calcul non-linéaire permettrait d'obtenir les états d'endommagement plus avancés de la structure.

## 7. Bibliographie